\documentclass[10pt,preprintnumbers,amsmath,amssymb,nofootinbib
,superscriptaddress]{revtex4}
\usepackage{graphicx,longtable,mathrsfs,color,array}
\usepackage[hidelinks]{hyperref}
\usepackage[usenames,dvipsnames]{xcolor} 
\usepackage{amssymb,amsmath,mathtools,mathrsfs,slashed} 
\usepackage{epsfig,subfigure,placeins,float} 
\usepackage{booktabs,longtable,ctable,multirow} 
\usepackage{exscale,relsize} 
\usepackage[normalem]{ulem} 
\usepackage{enumerate}
\usepackage{times, mathptmx} 
\usepackage[utf8]{inputenc}
\usepackage{color}
\usepackage{hyperref}
\usepackage{graphicx}
      \usepackage{color}
\usepackage{graphicx,graphics}
\usepackage{bm}
\usepackage{tabularx}
\allowdisplaybreaks[1]

\newcommand{\be}{\begin{equation}}  
\newcommand{\ee}{\end{equation}}  
\newcommand{\bea}{\begin{eqnarray}}  
\newcommand{\eea}{\end{eqnarray}}  
\newcommand{\ba}{\begin{array}}  
\newcommand{\ea}{\end{array}}

\newcommand{\jn}[1]{\textcolor{red}{[{\bf JN}: #1]}}

\newcommand{\comment}[1]{}

\newcommand{\MPl}{M_P}

\begin{document}

\title{$R^2$/Higgs inflation and the hierarchy problem}

\author{Pedro G. Ferreira}
\email{pedro.ferreira@physics.ox.ac.uk}
\affiliation{Astrophysics, University of Oxford, DWB, Keble Road, Oxford OX1 3RH, UK}
\author{Christopher T. Hill}
\email{hill@fnal.gov}
\affiliation{Fermi National Accelerator Laboratory, P.O.Box 500, Batavia, Illinois 60510, USA}
\author{Johannes Noller}
\email{johannes.noller@port.ac.uk}
\affiliation{Institute of Cosmology \& Gravitation, University of Portsmouth, Portsmouth, PO1 3FX, UK}
\affiliation{Department of Applied Mathematics and Theoretical Physics, University of Cambridge, Cambridge, CB3 0WA, UK}
\author{Graham G. Ross}
\email{g.ross1@physics.ox.ac.uk}
\affiliation{Theoretical Physics, University of Oxford, 1 Keble Road, Oxford OX1 3NP, UK}
\date{Received \today; published -- 00, 0000}

\begin{abstract}
We analyse Starobinsky inflation in the presence of the Brout Englert Higgs (BEH) boson with a non-minimal coupling  to the Ricci scalar, $R$. The latter  induces a coupling of the massive scaleron associated with the $R^2$ term to the BEH boson and this leads to a radiative correction to the BEH mass that must be fine tuned to keep the scalar light. For the case of $R^2$ driven inflation this requires a high level of fine tuning of order 1 part in $10^{8}$; for the case of Higgs inflation it is very much greater. We consider a scale invariant extension  of the $R^2$/Higgs model and find that for $R^2$ driven inflation but not for Higgs inflation the required fine tuning  is significantly reduced to one part in $10^{3-4}$. We consider the vacuum stability of the fine tuned model and its reheating and dilaton abundance after inflation. We also discuss possible gravitational wave signals associated with the model and the constraint on the mass of scalar or fermion dark matter candidates if they are produced by the gravitational couplings of the scalaron. 
  \end{abstract}

\maketitle
\section{Introduction}
Starobinsky's pioneering work on inflation led to his introduction of the Starobinsky model \cite{Starobinsky:1980te} based on an extension of General Relativity (GR) to include the $R^2$ curvature squared corrections to the Einstein Hilbert action that are expected to occur in radiative order.  Subsequently it was realised that when the BEH boson \cite{Englert:1964et,Higgs:1964ia} of the Standard Model (SM) of the strong, weak and electromagnetic interactions  is included the ``$R^2$/Higgs" model can also lead to ''Higgs" inflation \cite{Bezrukov:2007ep}. It is remarkable that the  model successfully predicts the inflationary parameters \cite{Mukhanov:1981xt} observed by Planck \cite{Aghanim:2018eyx,Akrami:2018odb}   and is entirely consistent with current observations. Here we consider a potential problem of the model which requires the co-existence of two vastly different mass scales, the Planck scale, $M_P=2.44\times10^{18}$ {\rm GeV}, and the electroweak scale characterised by the BEH vacuum expectation value, $v=246$ {\rm GeV}. This introduces a ``naturalness" issue for it is considered to be unnatural to have such different scales as radiative corrections typically drive the scales together.

In the absence of states with mass much larger than the electroweak scale the radiative corrections to the BEH mass of order the Planck scale only depend on the ultraviolet (UV) completion of the model which may require the absence of such corrections  (for example if the UV theory is scale invariant).  It is in this sense that the SM together with the Einstein Hilbert action describing gravity can be considered natural as it does not require the existence of particles with mass much larger than the electroweak scale \cite{Bardeen:1995kv}. However the Starobinsky model {\it does} require the appearance of a new scalar degree of freedom, the ``scaleron", that has non-minimal coupling to the Ricci scalar and a mass related to the Planck scale. Through its non-minimal coupling it can introduce significant corrections to the BEH mass leading to an unacceptable hierarchy problem and it is this aspect that we consider in detail here. 

As emphasised by Bardeen \cite{Bardeen:1995kv}, the  dangerous terms are those that vary logarithmically with the scale at which they are measured for they cannot be eliminated by the symmetry of the underlying UV theory. As we will discuss a calculation of the radiative corrections to the BEH mass in the $R^2$/Higgs model shows that the fine tuning needed to keep the BEH scalar light requires extreme fine tuning making the model problematic. However we find that, for the case when inflation is driven by the $R^2$ term,  a scale invariant extension of the model, introduced to tame its power law divergences,  significantly reduces the logarithmically varying correction which, while still large, may be acceptable. For the fine tuned version of the scale invariant model we consider the implications of demanding vacuum stability during inflation, the reheat phase and the dilaton abundance today, the possibility of testing the model via gravitational wave measurements and the nature of additional states capable of describing the observed dark matter abundance.

In Section \ref{Star} we introduce the scale dependent $R^2$/Higgs model and the parameter range leading to $R^2$ or Higgs inflation. In Section \ref{scaleinv} we define the scale invariant version of the model and determine the relation to the scale dependent case showing how the inflationary parameters are related. Section \ref{FT} presents the fine tuning for both the scale dependent and scale invariant versions. We then consider various phenomenological issues for the scale invariant version constrained to arrange its necessary fine tuning. In Section \ref{VS} we discuss its vacuum stability during inflation; in Section \ref{RH} we determine the reheating mechanism; in Section \ref{DA} we determine the resultant dilaton abundance; in Section \ref{BH} we consider the prospect for observing gravitational wave signals resulting from Black Hole formation and in Section \ref{DM} we consider the possibility of extending the model further to obtain the observed dark matter abundance.
\section{The mixed $R^2$/Higgs model}\label{Star}
In the presence of the BEH scalar, the Starobinsky model is given by the action
\bea
\label{R2S}
S &=& \int\sqrt{-g}\left(\frac{1}{2}g^{\mu\upsilon}\partial_{\mu}{H} \partial_{\nu
}{H}-\frac{1}{6f_{0}^{2}}R^{2}-\frac{1}{12}\alpha_H 
H^2 R+\frac{1}{2}%
M_P^{2}R-\frac{\lambda}{4}(H^2-v^2)^2 \right)
\eea
where $M_P$ is the Planck mass and ${R}({g})$ is the Ricci scalar of the metric $g_{\alpha\beta}$.  
$H$ is a real scalar field, a component of the  BEH isodoublet scalar, $\mathcal{H}$, of the Standard Model given, in the unitary gauge where the electroweak gauge fields are pure-gauge configurations, by $\mathcal{H}=(0,H/\sqrt{2})$. 

Since the $R^2$ term involves fourth order derivatives it contains an additional (scalar) degree of freedom \cite{Whitt:1984pd,Hindawi:1995an} 
.  To make this explicit it is convenient to reduce the fourth order derivatives to second order by introducing the
auxiliary ''scaleron" field $\eta$  with the action now given by 
\bea
S=\int\sqrt{-g}\left(\frac{1}{2}g^{\mu\upsilon}\partial_{\mu}%
H\partial_{\nu}H+\frac{1}{2}M_P^{2}R\Omega^{2}\ \ -\frac{\xi}{4}\eta^{4}-\frac{\lambda}{4}(H^2-v^2)^2
\label{NSI1}
\right)
\eea
where 
\bea
\xi&=&f_{0}^2\alpha_\eta^2/24\nonumber\\
\Omega^{2}&=&  1-\frac{\alpha_{H}}{6M_P^{2}}%
H^{2}-\frac{\alpha_\eta}{6M_P^{2}}\eta^{2} \equiv\exp\left(  \frac{2\chi
}{\sqrt{6}M_P}\right)  
\label{NSI2}
\eea
and $\chi$ plays the role of the inflaton in the case of $R^2$ dominated inflation.

As discussed in  \cite{Hill:2020oaj,Hill:2021psc} the anomalous coupling terms to the Ricci scalar  proportional to $\alpha_H$ and $\alpha_\eta$ lead to contact terms in the effective Lagrangian below the Planck scale driven by single graviton exchange where the $\frac{1}{Q^2}$ terms coming from the graviton propagator are cancelled by $Q^2$ terms in the numerator coming from vertex factors. The resulting $Q^2$ independent local ``contact" terms  must be included as pointlike interactions in the effective Lagrangian. The contact terms are the same as are usually obtained from a Weyl transformation going from the ``Jordan" frame to the ``Einstein" frame\footnote{Except that the metric is not transformed.} and show that the Jordan frame is already the Einstein frame in the effective Lagrangian. Including these contact terms the action has the form 
\bea
S=\int\sqrt{-g}\left(\frac{1}{2}\Omega^{-2}g^{\mu\upsilon}\partial_{\mu}%
H\partial_{\nu}H+\frac{1}{2}g^{\mu\upsilon}\partial_{\mu}%
\chi\partial_{\nu}\chi+\frac{1}{2}M_P^{2}R\ \ -\frac{\xi}{4}\Omega^{-4}\eta^{4}-\frac{\lambda}{4}\Omega^{-4}(H^2-v^2)^2
\label{NSIF}
\right)
\eea
where
\bea
\label{pot17}
\frac{\xi}{4}\Omega^{-4}\eta^{4}&=&\frac{3}{8}M_P^{4}f_{0}^{2}
\left(
1-\exp\left( -\frac{2 \chi }{\sqrt{6}M_P}\right)
\left(  1-\frac{\alpha_{H}}{6M_P^{2}}H^{2}\right)  \right)  ^{2}.
\eea

Note that even if the anomalous coupling, $\alpha_H$,  is initially zero, it is generated radiatively by Standard Model couplings. Its beta function, $\beta_\alpha$, scales as the beta function for the BEH mass \cite{Buchbinder:1992rb}, and has the form \cite{Herranen:2014cua}
\be
16{\pi ^2}{\beta _{\alpha_H} } = \left( {\alpha_H  -1} \right)\left( {12\lambda  + 6y_t^2 - \frac{3}{2}g{'^2} - \frac{9}{2}{g^2}} \right)
\label{BETA}
\ee
The origin of the $ \left( {\alpha_H  -1} \right)$ factor is that at the conformal symmetry point the couplings generating the radiative corrections are scale invariant and and so the beta functions is only generated by perturbations away from this point.

\subsection{$R^2$ or Higgs inflation}
As discussed in \cite{He:2020ivk,Enckell:2018uic,Gundhi:2018wyz} for a large initial $\phi$ VEV and sufficiently large $\lambda$, there is an attractor behaviour that causes the scalar fields to fall in to one of the two valleys in the two field potential and then there is slow roll inflation corresponding to ``Higgs" or ``$R^2$" inflation with the dominant term during inflation being driven by the BEH scalar or the scaleron respectively. 
Using the relation between $\eta$ and $H$ in the valley, inflation effectively becomes single field inflation and can be studied analytically. Inflation is in the valley corresponding to Starobinsky inflation driven by the $R^2$ term in the region \cite{He:2020ivk}
\be
0\le \alpha_H\le \alpha_c/\sqrt{2},\;\;\;\alpha_c=2.6\times10^4\sqrt{\lambda/0.01}
\ee
while  in the region 
\be
\alpha_c/\sqrt{2}\le \alpha_H\le \alpha_c.
\ee
the scalar fields are driven to the second valley corresponding to Higgs inflation. When $\alpha_H>\alpha_c$ the system becomes nonperturbative.

To leading order in $\frac{1}{N_*}$ (where $N_*$ is the number of e-foldings\footnote{Here $N_\star$  refers to e-foldings in Einstein or Jordan frame -- once contact terms are added they are the same.}), the scalar spectral index and tensor to scalar index for both regimes are independent of the parameters and both Higgs and Starobinsky inflation are in the same universal attractor regime giving $0.939<n_R<0.967$ for the spectral index and $3.8\times 10^{-3}<r<0.079$ for the tensor to scalar ratio \cite{Enckell:2018uic}
in excellent agreement with the Planck data \cite{Akrami:2018odb}.   At this order the only constraint on model parameters comes from the normalisation of the scalar amplitude of the primordial curvature fluctuations which gives \cite{Gundhi:2018wyz}
\be
\frac{\lambda}{\alpha_H^2+24\lambda/f_0^2 }\approx 1.2\times 10^{-11}.
\ee 

For $R^2$ inflation this requires $f_0=1.7\times 10^{-5}$.
Higgs inflation corresponds to the region $f_0^2\gg\tfrac{24\lambda}{\alpha_H^2}$. There is an upper bound on $f_0$ given by   $f_0^2\leq\frac{4\lambda}{\alpha_H}$ \cite{Enckell:2018uic} coming from the fact that higher order corrections to the inflationary parameters diverge in the limit $\frac{\lambda}{\alpha_Hf_0^2}\rightarrow 0$.
Requiring the observed scalar amplitude fluctuations then gives $\alpha_H\approx 2\times 10^4$ \cite{He:2020ivk} corresponding to $10^{-3}\gtrsim f_0\gg 10^{-5}$ where we have taken $\lambda=0.01$ at the scaleron mass scale.
The lower limit is close to the non-perturbative limit \cite{Ema:2017rqn,Gorbunov:2018llf},$f_0\le \frac{\sqrt{24\pi}}{\alpha_H}\approx 10^{-3}.$
The inflationary observables predicted by $R^2$ and Higgs inflation are very similar; for a discussion of the differences see \cite{Bezrukov:2011gp}. 

After inflation the scaleron mass is $M_Pf_0/\sqrt{2}$. For  Starobinsky inflation the normalisation of the scalar amplitude requires $f_0\approx 10^{-5}$ corresponding to a scaleron mass of $O(10^{13}{\rm GeV})$. For Higgs inflation the mass is larger.

\section{A scale invariant version of the mixed $R^2$/Higgs model}\label{scaleinv}
To obtain a scale invariant version of the Starobinsky model we generate the Planck scale via a new scalar field, $\phi$, that spontaneously breaks the scale invariance when it acquires a Planck scale vev \cite{ShapoZen,ShapoZen2,ShapoBlas,GarciaBellido:2011de,Ferreira:2016vsc,Ferreira:2016wem}.
The action is given by eq(\ref{R2S}) with the replacement 
\be
{M_P^2\over 2}\;R({\tilde g})\rightarrow  {1\over 2}g^{\mu\nu}\partial_\mu\phi\partial_\nu\phi 
-\frac{\tau}{4}\phi^4-{1\over 12}\alpha_\phi\phi^2 R({\tilde g})
\label{R2Sinv}
\ee
Proceeding as in Section \ref{Star} by introducing  the auxiliary field $\eta$, which can be integrated out to recover the $R^{2}$ term we obtain the scale Invariant Starobinsky action in the form:
\bea
S=\int\sqrt{-g}\left(\frac{1}{2}g^{\mu\upsilon}\partial_{\mu}\phi
\partial_{\nu}\phi+\frac{1}{2}g^{\mu\upsilon}\partial_{\mu}%
H\partial_{\nu}H+\frac{1}{2}M_P^{2}R\Omega^{2}\ \ -\frac{\xi}{4}\eta^{4}-\frac{\omega}{4}\phi^{4}-\frac{\lambda}{4}H^{4})
\right)
\label{a1I}
\eea
where 
\bea \label{a2prev}
\Omega^{2}=  -\frac{\alpha_{\phi}}{6M_P^{2}}%
\phi^{2}-\frac{\alpha_{H}}{6M_P^{2}}%
H^{2}-\frac{\alpha_\eta}{6M_P^{2}}\eta^{2}\eea

As discussed in \cite{GarciaBellido:2011de,Ferreira:2016wem,Ghilencea:2018thl}
this theory is invariant under a global Weyl (scale) transformation.
Due to the conservation of the Weyl current associated with the scale symmetry the system undergoes inertial spontaneous scale symmetry breaking \cite{Ferreira:2018itt} with the appearance of a massless dilaton, $\sigma(x)$. It is convenient to change variables to 
\be \label{dilatonDecouplingTrans}
g_{\mu\nu}=e^{2\sigma(x)/f}\hat g_{\mu\nu},\;\;\Theta_i=e^{-\sigma(x)/f}\hat \Theta_i
\ee
where f has dimensions of mass and $\Theta_i$ runs over the three scalar fields. The dilaton, $\sigma$, apparently decouples in the Jordan frame \cite{Ferreira:2016kxi} but acquires derivative couplings to Standard  Model fields via contact terms \cite{Hill:2021psc}.

The conserved Noether current, $K_\mu$ can be written in terms of a kernel $K$:
\bea
K_{\mu}=\partial_{\mu}K, \qquad
K=\frac{1}{6}\left(  (1-\alpha_\phi)\widehat{\phi}^{2}+(1-\alpha_H%
)\widehat{H}^{2}-\alpha_\eta\widehat{\eta}^{2}\right)   e^{-2\sigma/f}\equiv \bar{K} e^{-2\sigma/f}
\label{kernal}
\eea
where $\bar{K}$ is a constant.
Inertial symmetry breaking  \cite{Ferreira:2018itt} can be viewed as the red shifting of the dilaton
to a constant VEV, $\sigma\rightarrow \sigma_0+\hat\sigma(x)$,  and thus
$K\rightarrow \overline{K}e^{-2\sigma_0/f}\equiv M_P^2$  giving 
\bea
S=\int\sqrt{-g}\left(\frac{1}{2}\partial_{\mu}\hat\sigma
\partial^{\mu}\hat\sigma +\frac{1}{2}\partial_{\mu}\hat\phi
\partial^{\mu}\hat\phi+\frac{1}{2}\partial_{\mu}
\hat H\partial^{\mu}\hat H+\frac{1}{2}M_P^{2}R\hat\Omega^{2}\ \ -\frac{\xi}{4}\hat\eta^{4}-\frac{\omega}{4}\hat\phi^{4}-\frac{\lambda}{4}\hat H^{4})
\right)
\label{a12}
\eea

where, using eq(\ref{kernal}), 
\be
\hat\Omega^2=1-\frac{1}{6M_P^{2}}%
\hat\phi^{2}-\frac{1}{6M_P^{2}}\hat H^{2}\equiv\exp\left(  \frac{2\theta
}{\sqrt{6}M_P}\right).  
\label{a2}
\ee
Finally, replacing the anomalous couplings to the Ricci scalar by the contact terms in the effective Lagrangian (as discussed above this is equivalent to a Weyl transformation \cite{Hill:2020oaj}), gives the final form
\bea
S=\int\sqrt{-g}\left(\frac{1}{2}M_P^{2}R+\frac{1}{2}\partial_\mu\theta\partial^\mu\theta+\frac{1}{2}\hat\Omega^{-2}\left(\partial_{\mu}\hat\sigma
\partial^{\mu}\hat\sigma +\partial_{\mu}\hat\phi
\partial{\mu}\hat\phi+\partial_{\mu}
\hat H\partial^{\mu}\hat H\right) -\frac{\xi}{4}\hat\Omega^{-4}\hat\eta^{4}-\frac{\omega}{4}\hat\Omega^{-4}\hat\phi^{4}-\frac{\lambda}{4}\hat\Omega^{-4}\hat H^{4})
\right)
\label{a1}
\eea
where, from eq(\ref{kernal}), the potential coming from the ${\xi\Omega^{-4}\over4}\eta^4$ term is
\be
V={3\over 8}M_P^4f_0^2\left(1+{{\alpha_\phi\over 6} {\hat\phi^2\over M_P^2}+{\alpha_H\over 6}{\hat H^2\over M_P^2}\over 1-{1\over 6}{\hat\phi^2\over M_P^2}-{1\over 6}{\hat H^2\over M_P^2}}\right)^2
\label{SIpotential}
\ee

Note that there is an extended SO(2) symmetry of the action in the limit \ $a_\phi=a_H$, \ $\lambda
= \omega$. There is also a $Z_2$ symmetry under which only the dilaton is odd and so it can only be produced in pairs.

\subsection{Effective single field dynamics and inflationary observables}

In order to compare the inflationary predictions for the scale dependent and scale independent models  it is useful to re-cast these two models in terms of effective single scalar field theories.  For simplicity we will consider only the $R^2$ inflation regime and ignore the BEH scalar in this context; it is straightforward, if algebraically messy, to generalise the discussion to include the BEH scalar.
When introducing an auxiliary field as in \eqref{NSI1} and replacing the anomalous couplings by the associated counter terms in the effective Lagrangian, the original Starobinsky inflation model can be written as 
\begin{align}
S_{R^2} = \MPl^2 \int d^4x \sqrt{-g}\Big[\frac{R}{2}
-  \frac{3 \alpha_\eta^2 \eta^2}{(6 M_{\rm Pl}^2 -  \alpha_\eta \eta^2)^2}\partial_{\mu}\eta \partial^{\mu}\eta
- \frac{9 M_{\rm Pl}^2 \xi \eta^4}{(6 M_{\rm Pl}^2 -  \alpha_\eta \eta^2)^2} 
\Big].
\label{StaroSTEinsteinFrame}
\end{align}
Canonically normalising the scalar field by sending $\alpha_\eta\eta^2 \to 6 M_P^2(1 - e^{\frac{\sqrt{\frac{2}{3}}\pi}{M_P}})$, we can finally write this as 
\begin{align}
S_{R^2} = \int d^4x \sqrt{-g} \Big[ \frac{\MPl^2}{2} R -  \frac{1}{2} \partial_{\mu}\pi \partial^{\mu}\pi - V_{R^2}(\pi)\Big], \quad\quad \text{where} \quad\quad 
V_{R^2}(\pi) = \frac{3}{8}\MPl^4 f_0^2  e^{-2\sqrt{\frac{2}{3}}\frac{\pi}{\MPl}}\left(e^{\sqrt{\frac{2}{3}}\frac{\pi}{\MPl}}-1\right)^2.
\label{StaroSTfinal}
\end{align}
%
%
This gives us an effective single field version of the original Starobinsky model, canonically normalised and in the Einstein frame. 

Consider now the case of the scale invariant version given in eq(\ref{a1}). As for the scale dependent case we simplify the analysis by ignoring the BEH scalar and also the dilaton. Using eq(\ref{a2}) we have
\be 
(\partial\theta)^2=\frac{\frac{\phi^2}{6M^2}}{(1-\frac{\phi^2}{6M^2})^2}(\partial\phi)^2
\ee
giving the kinetic term
\bea
\frac{1}{2}(\partial\theta)^2+\frac{1}{2}\Omega^{-2}(\partial\phi)^2&=&\frac{1}{2}\frac{1}{(1-\frac{\phi^2}{6M^2})^2}(\partial\phi)^2
\eea
This can be rewritten in canonical form by changing the variable to
\be
\chi=\sqrt{\frac{3}{2}}M_P\log\left(\frac{\sqrt{6}+\frac{\phi}{M_P}}{\sqrt{6}-\frac{\phi}{M_P}}\right),\;\;i.e.\;\;
\phi=\sqrt{6}M_P\frac{(e^{\sqrt{\frac{2}{3}}\frac{\chi}{M_P}}-1)}{(e^{\sqrt{\frac{2}{3}}\frac{\chi}{M_P}}+1)}= \sqrt{6}\MPl{\rm Tanh}\left[\frac{\chi}{\sqrt{6}\MPl}\right].
\ee
Using this in eq(17) gives the $\chi$ dependent potential in the form
\be
V(\chi)=\frac{3}{8}M_P^4f_0^2\left(1+\frac{\alpha_\phi(e^{\sqrt{\frac{2}{3}}\frac{\chi}{M_P}}-1)^2}{4e^{\sqrt{\frac{2}{3}}\frac{\chi}{M_P}}}\right)^2 =
\frac{3}{32} \MPl^4 f_0^2 \left(2 - \alpha_\phi + \alpha_\phi {\rm Cosh}\left[\sqrt{\tfrac 23} \tfrac \chi\MPl\right]\right)^2.
\ee
Inflation ends when $\chi=\chi_0$, where
\be
e^{-2\sqrt{\frac{2}{3}}\frac{\chi_0}{M_P}}\approx -\frac{\alpha_\phi}{2}
\ee

Thus, to analyse the last stage of inflation, it is convenient to expand $\chi$ in the neighbourhood of the inflation end point, $\chi=\chi_o-\tilde\chi$.
This gives 
\be
V(\chi)\approx\frac{3}{8}M_P^4f_0^2\left(1-\sqrt{\frac{\alpha_\phi}{2}}e^{-\sqrt{\frac{2}{3}}\frac{\chi}{M_P}}+O(\alpha_\phi)\right)^2 =\frac{3}{8}M_P^4f_0^2\left(1-e^{-\sqrt{\frac{2}{3}}\frac{\tilde\chi}{M_P}}+O(\alpha_\phi)\right)^2
\ee
where $\tilde\chi=\chi+\sqrt{\frac{3}{2}}\log(\sqrt{\frac{\alpha_\phi}{2}})$.

Up to terms suppressed by powers of the small parameter, $\alpha_\phi$, this is the same as the potential for the scale dependent case given in eq(24) showing that, the inflationary predictions of the scale dependent and scale independent theories are approximately the same.

\section{Fine tuning in the $R^2$/Higgs model and its phenomenological implications.}

\subsection{Fine tuning}\label{FT}
In order to quantify the amount of fine tuning needed for consistency with the observed BEH mass it is necessary to calculate the correction to the mass resulting from its coupling to the massive scaleron either by directly computing the effective potential via the Coleman Weinberg method or using the Renormalisation Group approach. After including the contact terms this has been done for both the scale dependent and scale invariant $R^2$/Higgs models \cite{Hill:2021psc} and here we just quote the results.

\subsubsection{The scale dependent $R^2$/Higgs model}

The one-loop induced mass term has the form  \bea
\Gamma_H  =-\frac{1}{768\pi^{2}}M_P^{2}f_{0}^{4}\alpha_{H}\left(
\alpha_{H}-3\right)  h^{2}\left(  \ln\left( M_P^{2}/m^{2}\right)
\right)  
\label{mt1SD}
\eea
Note that it does not vanish in the conformal limit $\alpha_H=1$. 

For Higgs inflation $\alpha_H\approx 3\times 10^3$ and $f_0\gg10^{-5}$ implying that $\delta m_H$, the radiative correction to the Higgs mass, is given by $\delta m_H\gg 10^{13} {\rm GeV}$. Consistency with the observed Higgs mass then requires an extremely unnatural cancellation with the bare mass term with accuracy much greater than 1 part in $10^{13}$. Since this correction varies logarithmically below the Planck scale the cancellation cannot be due to the underlying UV symmetry or dynamics at the Planck scale.

For inflation driven by the $R^2$ term the normalisation to the observed scalar perturbations requires $f_0\approx 10^{-5}$. This corresponds to a typical contribution\footnote{Taking $m_H^2=0$ at the Planck scale} to the BEH mass squared of $O( 10^{13} \alpha_H(\alpha_H-3){\rm GeV}^2)$. Thus, to get a contribution no greater than the observed BEH mass, $\alpha_H$ evaluated at the scaleron mass scale must be no greater than $10^{-9}$. From eq(\ref{BETA}) we see that the radiative corrections to $\alpha_H$ from SM states are typically of $O(0.1)$ so the bound on $\alpha_H$ implies a fine tuning in the initial value of $\alpha_H$ of order 1 part in $10^{8}$, smaller but still unacceptably large.

\subsubsection{The scale independent $R^2$/Higgs model}
The one-loop induced mass term has the form  
\bea
\Gamma_H  =-\frac{1}{768\pi^{2}}M_P^{2}f_{0}^{4}\alpha_{H}\left(
\alpha_{H}-\alpha_\phi\right)  (1-\alpha_\phi)h^{2}\left(  \ln\left( M_P^{2}/m^{2}\right)
\right)  
\label{mt1}
\eea
which also does not vanish in the conformal limit $\alpha_H=1$.

As mentioned above, to get viable inflation it is necessary to have small $\alpha_\phi$ and indeed the inflationary predictions are consistent with observation for $\alpha_\phi=0$ \cite{Ferreira:2018qss,Ferreira:2019zzx}. In this case 
\be
\beta_{m_H^2}=-{M_P^2\over 192\pi^2}f_0^4 \alpha_H^2
\ee

For the case of Higgs inflation the situation is the same as for the scale dependent case requiring a fine tuning greater than 1 part in $10^{13}$.

For the case of $R^2$ dominated inflation, $f_0\approx 10^{-5}$ and, taking $m_H^2=0$ at the Planck scale,  leads to a BEH mass squared of $O(10^{13}\alpha_H^2{\rm GeV}^2)$ corresponding to a required fine tuning in $\alpha_H$ of 1 part in $10^{3-4}$, significantly reduced compared to the scale dependent case.  
\subsection{Vacuum stability}\label{VS}

For the central values of the Standard Model parameters the BEH potential has a second deeper minimum at a scale above $10^{10}{\rm GeV}$ where the coefficient of the quartic BEH term, $\lambda$, becomes negative \cite{Degrassi:2012ry,Buttazzo:2013uya,Isidori:2001bm,Sher:1993mf,Ellis:2009tp,Casas:1996aq,EliasMiro:2011aa}.
 The radiative corrections due to the scaleron can be relevant \cite{Avramidi:1986mj,Gorbunov:2018llf} but these are negligible for the case $\alpha_H\approx 0$ and in any case they only affect the evolution above the scaleron mass. 
 
 For the case of Higgs inflation it is necessary that the quartic coupling remain positive up to the Planck scale so one must assume the top mass is 3$\sigma$ lighter than its world average central value or $2\sigma$ lighter than the recent CMS measurement \cite{1979577}. 
 
 For the case of $R^2$ inflation it is possible to live with the second minimum provided one is driven to the unstable but long-lived minimum at the electroweak scale.
 The requirement of false vacuum stability  puts bounds on the BEH non-minimal coupling. 
 \cite{Markkanen:2018pdo,Buchbinder:1992rb,Markkanen:2018bfx,Rajantie:2017ajw,Herranen:2015ima,Figueroa:2017slm,Herranen:2014cua}.
  In particular a lower bound  $\alpha_H>0.23$ at the scaleron mass scale is needed to overcome the second minimum during inflation, inconsistent with the value $\alpha_H\lesssim 10^{-4}$ needed to avoid the hierarchy problem. Thus, as for the Higgs inflation case, for consistency during inflation it is necessary that the BEH quartic coupling is not driven negative which in turn requires that the the top quark should  be some 2-3 $\sigma$ lighter than the current best fit value. 
\subsection{Reheating}\label{RH}
We are interested in what happens at the end of inflation when, up to electroweak scale corrections,  $\eta=0$. 
Reverting to the $\phi$, $H$ basis, and ignoring the $H$ v.e.v. this corresponds to the $\phi$ v.e.v., $\phi_0$, given by
\be
\phi_0^2={6 M_P^2\over 1-\alpha_\phi}.
\ee 
Reheating in the Starobinsky model has been studied by several authors \cite{Starobinsky:1980te},\cite{Vilenkin:1985md}-\cite{Bernal:2020qyu}. Here we calculate the reheat temperature for the scale invariant model introduced in Section \ref{scaleinv}. 
For small fluctuations about the minimum given by  $\phi=\phi_0+\hat\phi$ the  fields, $\phi_N$, $H_N$, that give the correctly normalised  kinetic term follows from eq(\ref{a1}) and are given by
\be
\hat\phi={-\alpha_\phi\over \sqrt{1-\alpha_\phi}}\phi_N,\;\;\;H=\sqrt{\frac{-\alpha_\phi}{1-\alpha_\phi}}H_N\;\;\;\sigma=\sqrt{\frac{-\alpha_\phi}{1-\alpha_\phi}}\sigma_N
\label{norm}
\ee
so the mass, $m_{\phi}$, of the normalised field, $\phi_N$, is $(1-\alpha_\phi)^{3/4}\frac{f_0}{\sqrt{2}}M_P\sim10^{13}{\rm GeV}$. 
The dominant interactions of the scaleron are to the BEH scalar and the dilaton as the decay rate to fermions is suppressed by the factor $(\frac{m_f}{M_P})^2$  \cite{Gorbunov:2010bn}.

There are  interaction terms coming from the potential, eq(\ref{SIpotential}) and expressing the result in terms of the normalised fields gives the leading order  terms
\be
(1-\alpha_H)(1-\alpha_\phi)\frac{f_0^2}{4\sqrt{6}}M_P\phi_NH_NH_N
\label{rh2}
\ee
There are also  interaction terms associated with the kinetic terms of eq(\ref{a1}) given by
\be
\frac{\phi_N}{\sqrt{6}M_P}\left(\partial^\mu\sigma_N\partial_\mu\sigma_N +\partial_\mu H_N\partial^\mu H_N\right)\sim\frac{m_\phi^2}{2\sqrt{6}M_P}(\phi_NH_NH_N+\phi_N\sigma_N\sigma_N)=(1-\alpha_\phi)^{3/2}\frac{f_0^2}{4\sqrt{6}}M_P(\phi_NH_NH_N+\phi_N\sigma_N\sigma_N)
\label{rh1}
\ee
where, when calculating the decay width, the derivative terms may be expressed in terms of the scaleron mass.
From eqs(\ref{rh1},\ref{rh2}), for negligible $\alpha_{H,\phi}$, the scaleron decay rate is of order
\be
\Gamma_{\phi\rightarrow HH}=4\Gamma_{\phi\rightarrow\sigma\sigma}=\frac{m_\phi^3}{ 96\pi M_P^2}
\label{scalarrate}
\ee
The reheat temperature is given by
\footnote{Note that this does not vanish in the conformal limit so the conformal anomaly contributions are sub-dominant. This differs from the behaviour assumed in \cite{Gorbunov:2012ns}.} \be
T_{reh}=0.1 g_*^{-1/4}({N_s m^3\over M_P})^{1/2}
\ee
where $N_S$ is the number of scalar components. For the BEH states $N_S=4$ and the number of relativistic species on decay being given by the Standard Model value, $g_*=106.75$, gives
\be
T_{reh}\approx 6\times 10^{9}{\rm GeV}.
\ee

\subsection{Dilaton abundance}\label{DA}
Following from eq(\ref{scalarrate}) we see that after reheat
 the initial abundance of the dilaton will be $1/4$ that of the Higgs and the other Standard Model particles. The dilaton is never in thermal equilibrium so in tracking its relative abundance to neutrinos one has to take account of the reheating of the neutrinos due to SM decays that occurs before the neutrinos decouple. Initially $N(m_\chi)$, the effective number of relativistic species at the scale $m_\chi$ is 427/4  and at neutrino decoupling $N(T_{dec})=43/4$ so the relation of the initial neutrino temperature to that at neutrino decoupling is given by $T_\nu(m_\chi)=(43/427)^{1/3}T_\nu(T_{dec})$. The dilaton has one degree of freedom compared to 2 for a neutrino so its contribution to $N_{eff,\nu}$ today is 
\be
N_{eff,\chi}=\frac{1}{4}\times \frac{8}{7}\times\frac{1}{2}\times(\frac{43}{427})^{\frac{4}{3}}=0.007
\ee
With a detailed analysis of neutrino decoupling one has $N_{eff,\nu}=3.046$ for 3 active neutrinos so with the dilaton this gives $N_{eff,\nu + dilaton}=3.053$.
Current measurement of the Planck collaboration give measurement between  $N_{eff,\nu}=2.92\pm0.36$  and  $N_{eff}=3.27\pm0.15$ depending on what data sets are included \cite{Planck:2018vyg}. It does not seem possible to reduce the errors in the measurement to be sensitive to the dilaton contribution in the near future.

\subsection{Black Holes}\label{BH}
Starobinsky inflation belongs to the class of $f(R)$ models of modified gravity which, as we have seen, can be re-expressed as scalar-tensor theories \cite{Sotiriou}. As such,  vacuum black hole solutions are those of General Relativity (Schwarzschild and Kerr-Newman). It is possible, nevertheless, that perturbations of these black holes might lead to signatures which may distinguish them from General Relativity. Indeed, as pointed out in \cite{BarausseSotiriou} and \cite{TattersallFerreira}, the extra degrees of freedom (in this case, the scalaron) can be excited during a gravitational collapse event that leads to the formation of the black hole, affecting the gravitational wave signal through the non-minimal coupling. Intriguingly, this is not the case with Starobinsky inflation. At late times, the only vacuum solution (in the absence of an explicit cosmological constant) is Schwarzschild or Kerr-Newman and not, for example, Schwarzschild-DeSitter (we note that during inflation there is self-acceleration but this is transient). 
As such, the background value of the scalar field $\eta$ is zero and there is no mixing between the scalar fluctuations and the gravitational wave sector during, for example, ringdown.

This is not the case with the scale invariant Starobinsky model \cite{FerreiraTattersall}. There we have seen that there are two modes excited -- the Goldstone mode and a massive mode -- which will excite the gravitational wave spectrum. Thus, and in principle, there is a distinctive signature in black hole ring down which would distinguish between the scale invariant and ordinary Starobinsky mode. In practice, however, the mass scale associated with the observable scalar mode is far too high to be detectable with current or future gravitational wave detectors (see discussion in \cite{FerreiraTattersall}) .

\subsection{Dark Matter}\label{DM}
As there is no dark matter (DM) candidate in the globally scale invariant  Starobinsky model sector  it is of interest to ask what could be added to make DM. The situation is similar to that in the original Starobinsky model studied in \cite{Gorbunov:2010bn}. They consider the possibility it is due to new  fermionic or scalar states that only couple to Standard Model states through gravitational strength interactions.  For scalars the decay proceeds through the scaleron coupling to the kinetic term or through their contact terms and has the form of eq(\ref{scalarrate}). For fermions there is no such coupling to the kinetic term due to its conformally invariant structure and as a result the decay rate is suppressed 
\be
\Gamma_\phi\rightarrow \bar\psi\psi={m m_\psi^2\over 48\pi M_P}
\ee
Following [58] the masses of the new states needed to get the observed DM abundance is
\bea
m_\varphi\approx 6.9{\rm keV}\times({1.3\times 10^{-5}M_P\over m})^{1\over 2}({N_s\over 4})^{1\over 6}({g_*\over 106.75})^{1\over 4}({\Omega_{DM}\over 0.223})^{1\over 3}({\rho_c\over 0.52\times 10^{-5}{\rm GeV}/cm^3})^{1\over 3}\label{scalarDM}\\
m_\psi\approx 1.2\times 10^{7}{\rm GeV}\times({1.3\times 10^{-5}M_P\over m})^{-{1\over 2}}({N_s\over 4})^{1\over 6}({g_*\over 106.75})^{1\over 4}({\Omega_{DM}\over 0.223})^{1\over 3}({\rho_c\over 0.52\times 10^{-5}{\rm GeV}/cm^3})^{1\over 3}
\eea
where $N_s$ is the number of scalars contributing to inflaton decay and $\rho_c$ is the present energy density. However, since the states are highly relativistic on production with momentum some four orders of magnitude greater than the reheat temperature,  the new scalars give hot DM while the new fermions give cold DM. Thus the scalars can only provide a component of DM while the new fermions could provide all DM.

A   further possibility for DM arises if the scale symmetry is gauged and, on spontaneous breakdown, the would-be dilaton is ``eaten", providing  the longitudinal degree of freedom to the ``Weyl photon", $\omega_\mu$ \cite{Ghilencea:2018thl,Ferreira:2018itt,Ferreira:2016wem}. The gauging of the scale symmetry in the context of Weyl conformal geometry has the attractive feature that the spectrum is minimal, with no new fields beyond the Standard Model fields and those required by Weyl geometry \cite{Ghilencea:2021lpa} . 

The mass of the Weyl photon is proportional to $g_\omega M_{Planck}$  where $g_\omega$ is the gauge coupling associated with the scale symmetry. The most natural expectation is that it should have a mass  close to the Planck scale and be inflated away. Notwithstanding this it is of interest to ask whether, for anomalously small $g_\omega$, it could be DM. The Weyl photon couples to the scale current, $K_\mu=\partial_\mu \bar{K} e^{-2\sigma/f}$, given in eq(\ref{kernal}). Since $\bar K$ is constant the Weyl photon does not couple directly to the. scaleron   or the states of the Standard Model. Its coupling to the scaleron comes only from the contact term involving the photon, $\hat\Omega^{-2}F_{\mu\nu}F^{\mu\nu}$, analagous to the dilaton coupling in eq(\ref{a1}). Thus, as for the dilaton, the Weyl photon is the only state odd under a $Z_2$ symmerty. It is a absolutely stable and can only be produced in pairs. As such it is a  DM candidate.

If the Weyl photon is lighter than the scaleron its production is via scaleron decay through the scaleron coupling to the kinetic term. Thus the analysis of  Gorbunov and Panin \cite{Gorbunov:2010bn} applies and the mass of the Weyl photon is constrained to be at most of $O(keV)$.  Being hot DM it can only be a proportion of the observed DM abundance.

\section{Conclusions}

In this paper we have shown that the $R^2$/Higgs model of inflation suffers from a severe hierarchy problem. This arises because the massive scalaron associated with the $R^2$ term is coupled to the BEH boson and the resultant radiative corrections to the BEH mass are so large that, if it is to remain light, it is necessary to fine-tune couplings. For the case of the original  $R^2$/Higgs model the fine tuning required is  $1$ part in $10^{8}$ when inflation is driven by the $R^2$ term and very much worse for Higgs inflation.  This problem can be mitigated if we consider the scale-invariant version of the Starobinsky model in which the Planck mass is made dynamical and arises from inertial symmetry breaking. While the fine tuning remains unacceptably large for the case of Higgs inflation, for the case of $R^2$ driven inflation the fine tuning is much less severe albeit still problematic, of the order of $1$ part in $10^{(3-4)}$. 

The requirement of  vacuum stability during inflation plus the fine tuning constraint on the anomalous coupling of the BEH scalar to the Ricci scalar requires that the Standard Model potential should have no second minimum at a high scale and this in turn requires that the top mass should be some 2-3 $\sigma$ below its central value. Reheating in the model is dominantly through BEH scalar and dilaton production with a reheat temperature of $6\times 10^{9}{\rm GeV}.$ As the dilaton is never in thermal equilibrium its initial abundance is subsequently diluted, corresponding today only to an additional 0.007 of a neutrino.  Although the scaleron and the dilaton can excite the gravitational wave spectrum leading to distinctive black hole ring down structure, the signal is unobservable with current or future gravitational wave detectors.  

Finally, to explain the dark matter abundance it is necessary to extend the model to include a suitable candidate. For the case these states only interact gravitationally with the Standard Model states the DM mass is highly constrained. A scalar  should be light with mass  $\le O(7{\rm keV})$  and only contributes to the hot DM abundance. A fermion can be much heavier with mass  $ O(10^7 {\rm GeV})$ to provide the observed cold DM abundance. For the case the scale symmetry is gauged the dilaton is replaced by the massive Weyl photon. Its mass is naturally of order the Planck mass but can be much lighter if the scale symmetry gauge coupling is extremely small. If it is lighter than the scaleron,  DM constraints require that, similar to the scalar DM case, its mass should be $\le O(7{\rm keV})$ and it only contributes to the hot DM abundance.

\begin{acknowledgments}
We thank  O. Tattersall for helpful discussions. PGF acknowledges financial support from ERC Grant No 693024, the Beecroft Trust and STFC. JN is supported by an STFC Ernest Rutherford Fellowship, grant reference ST/S004572/1, and also acknowledges support from King's College Cambridge. This manuscript has been authored in part by Fermi Research Alliance, LLC under Contract No. DE-AC02-07CH11359 with the U.S. Department of Energy, Office of Science, Office of High Energy Physics.

\end{acknowledgments}


\begin{thebibliography}{99} 
   \bibitem{Starobinsky:1980te}   A.~A.~Starobinsky,   
  Phys.\ Lett.\ B {\bf 91} (1980) 99
   [Phys.\ Lett.\  {\bf 91B} (1980) 99]
   [Adv.\ Ser.\ Astrophys.\ Cosmol.\  {\bf 3} (1987) 130].
  
\bibitem{Englert:1964et}
F.~Englert and R.~Brout,
Phys. Rev. Lett. \textbf{13} (1964), 321-323
doi:10.1103/PhysRevLett.13.321

\bibitem{Higgs:1964ia}
P.~W.~Higgs,
Phys. Lett. \textbf{12} (1964), 132-133
doi:10.1016/0031-9163(64)91136-9, 
Phys. Rev. Lett. \textbf{13} (1964), 508-509
doi:10.1103/PhysRevLett.13.508

\bibitem{Bezrukov:2007ep}
F.~L.~Bezrukov and M.~Shaposhnikov,
Phys. Lett. B \textbf{659} (2008), 703-706
doi:10.1016/j.physletb.2007.11.072
[arXiv:0710.3755 [hep-th]].

\bibitem{Mukhanov:1981xt}
  V.~F.~Mukhanov and G.~V.~Chibisov,
  JETP Lett.\  {\bf 33} (1981) 532
   [Pisma Zh.\ Eksp.\ Teor.\ Fiz.\  {\bf 33} (1981) 549].]

\bibitem{Aghanim:2018eyx}
  N.~Aghanim {\it et al.} [Planck Collaboration],
  arXiv:1807.06209 [astro-ph.CO].
  
\bibitem{Akrami:2018odb}
  Y.~Akrami {\it et al.} [Planck Collaboration],
  arXiv:1807.06211 [astro-ph.CO].
  
  
\bibitem{Bardeen:1995kv}
W.~A.~Bardeen,
``On naturalness in the standard model,''
FERMILAB-CONF-95-391-T.
  

  \bibitem{Whitt:1984pd}   B.~Whitt,   
  Phys.\ Lett.\  {\bf 145B} (1984) 176.


  \bibitem{Hindawi:1995an}   A.~Hindawi, B.~A.~Ovrut and D.~Waldram,   
  Phys.\ Rev.\ D {\bf 53} (1996) 5583
  [hep-th/9509142].


\bibitem{Hill:2020oaj}
C.~T.~Hill and G.~G.~Ross,
Phys. Rev. D \textbf{102} (2020), 125014
doi:10.1103/PhysRevD.102.125014
[arXiv:2009.14782 [gr-qc]].

\bibitem{Hill:2021psc}
C.~T.~Hill and G.~G.~Ross,
[arXiv:2103.06827 [hep-th]].

\bibitem{He:2020ivk}
M.~He, R.~Jinno, K.~Kamada, A.~A.~Starobinsky and J.~Yokoyama,
JCAP \textbf{01} (2021), 066
doi:10.1088/1475-7516/2021/01/066
[arXiv:2007.10369 [hep-ph]].

\bibitem{Enckell:2018uic}
V.~M.~Enckell, K.~Enqvist, S.~Rasanen and L.~P.~Wahlman,
JCAP \textbf{01} (2020), 041
doi:10.1088/1475-7516/2020/01/041
[arXiv:1812.08754 [astro-ph.CO]].

\bibitem{Gundhi:2018wyz}
A.~Gundhi and C.~F.~Steinwachs,
Nucl. Phys. B \textbf{954} (2020), 114989
doi:10.1016/j.nuclphysb.2020.114989
[arXiv:1810.10546 [hep-th]].


\bibitem{Ema:2017rqn}
Y.~Ema,
Phys. Lett. B \textbf{770} (2017), 403-411
doi:10.1016/j.physletb.2017.04.060
[arXiv:1701.07665 [hep-ph]].

\bibitem{Gorbunov:2018llf}
D.~Gorbunov and A.~Tokareva,
Phys. Lett. B \textbf{788} (2019), 37-41
doi:10.1016/j.physletb.2018.11.015
[arXiv:1807.02392 [hep-ph]].

\bibitem{Bezrukov:2011gp}
F.~L.~Bezrukov and D.~S.~Gorbunov,
Phys. Lett. B \textbf{713} (2012), 365-368
doi:10.1016/j.physletb.2012.06.040
[arXiv:1111.4397 [hep-ph]].



  
\bibitem{Ferreira:2016wem}
P.~G.~Ferreira, C.~T.~Hill and G.~G.~Ross,
Phys. Rev. D \textbf{95} (2017) no.4, 043507
doi:10.1103/PhysRevD.95.043507
[arXiv:1610.09243 [hep-th]].
  
\bibitem{ShapoZen}
  M.~Shaposhnikov and D.~Zenhausern,
  Phys.\ Lett.\ B {\bf 671}, 162 (2009).


\bibitem{ShapoZen2}
  M.~Shaposhnikov and D.~Zenhausern,
  Phys.\ Lett.\ B {\bf 671}, 187 (2009)

\bibitem{ShapoBlas}
  D.~Blas, M.~Shaposhnikov and D.~Zenhausern,
  Phys.\ Rev.\ D {\bf 84}, 044001 (2011)


\bibitem{GarciaBellido:2011de}
J.~Garcia-Bellido, J.~Rubio, M.~Shaposhnikov and D.~Zenhausern,
Phys. Rev. D \textbf{84} (2011), 123504
doi:10.1103/PhysRevD.84.123504
[arXiv:1107.2163 [hep-ph]].

\bibitem{Ferreira:2016vsc}
P.~G.~Ferreira, C.~T.~Hill and G.~G.~Ross,
Phys. Lett. B \textbf{763} (2016), 174-178
doi:10.1016/j.physletb.2016.10.036
[arXiv:1603.05983 [hep-th]]. 



\bibitem{Ghilencea:2018thl}
D.~Ghilencea and H.~M.~Lee,
Phys. Rev. D \textbf{99} (2019) no.11, 115007
doi:10.1103/PhysRevD.99.115007

\bibitem{Ferreira:2018itt}
P.~G.~Ferreira, C.~T.~Hill and G.~G.~Ross,
Phys. Rev. D \textbf{98} (2018) no.11, 116012
doi:10.1103/PhysRevD.98.116012
[arXiv:1801.07676 [hep-th]].

\bibitem{Ferreira:2016kxi}
P.~G.~Ferreira, C.~T.~Hill and G.~G.~Ross,
Phys. Rev. D \textbf{95} (2017) no.6, 064038
doi:10.1103/PhysRevD.95.064038
[arXiv:1612.03157 [gr-qc]].

\bibitem{Ferreira:2018qss}
P.~G.~Ferreira, C.~T.~Hill, J.~Noller and G.~G.~Ross,
Phys. Rev. D \textbf{97} (2018) no.12, 123516
doi:10.1103/PhysRevD.97.123516
[arXiv:1802.06069 [astro-ph.CO]].

\bibitem{Ferreira:2019zzx}
P.~G.~Ferreira, C.~T.~Hill, J.~Noller and G.~G.~Ross,
Phys. Rev. D \textbf{100} (2019) no.12, 123516
doi:10.1103/PhysRevD.100.123516
[arXiv:1906.03415 [gr-qc]].

\bibitem{Ghilencea:2019rqj}
D.~Ghilencea,
JHEP \textbf{10} (2019), 209
doi:10.1007/JHEP10(2019)209
[arXiv:1906.11572 [gr-qc]].

\bibitem{Salvio:2017qkx}
A.~Salvio and A.~Strumia,
Eur. Phys. J. C \textbf{78} (2018) no.2, 124
doi:10.1140/epjc/s10052-018-5588-4
[arXiv:1705.03896 [hep-th]].

\bibitem{Degrassi:2012ry}
G.~Degrassi, S.~Di Vita, J.~Elias-Miro, J.~R.~Espinosa, G.~F.~Giudice, G.~Isidori and A.~Strumia,
JHEP \textbf{08} (2012), 098
doi:10.1007/JHEP08(2012)098
[arXiv:1205.6497 [hep-ph]].

\bibitem{Buttazzo:2013uya}
D.~Buttazzo, G.~Degrassi, P.~P.~Giardino, G.~F.~Giudice, F.~Sala, A.~Salvio and A.~Strumia,
JHEP \textbf{12} (2013), 089
doi:10.1007/JHEP12(2013)089
[arXiv:1307.3536 [hep-ph]].

\bibitem{Isidori:2001bm}
G.~Isidori, G.~Ridolfi and A.~Strumia,
Nucl. Phys. B \textbf{609} (2001), 387-409
doi:10.1016/S0550-3213(01)00302-9
[arXiv:hep-ph/0104016 [hep-ph]].

\bibitem{Sher:1993mf}
M.~Sher,
Phys. Lett. B \textbf{317} (1993), 159-163
doi:10.1016/0370-2693(93)91586-C
[arXiv:hep-ph/9307342 [hep-ph]].

\bibitem{Ellis:2009tp}
J.~Ellis, J.~R.~Espinosa, G.~F.~Giudice, A.~Hoecker and A.~Riotto,
Phys. Lett. B \textbf{679} (2009), 369-375
doi:10.1016/j.physletb.2009.07.054
[arXiv:0906.0954 [hep-ph]].

\bibitem{Casas:1996aq}
J.~A.~Casas, J.~R.~Espinosa and M.~Quiros,
Phys. Lett. B \textbf{382} (1996), 374-382
doi:10.1016/0370-2693(96)00682-X
[arXiv:hep-ph/9603227 [hep-ph]].

\bibitem{EliasMiro:2011aa}
J.~Elias-Miro, J.~R.~Espinosa, G.~F.~Giudice, G.~Isidori, A.~Riotto and A.~Strumia,
Phys. Lett. B \textbf{709} (2012), 222-228
doi:10.1016/j.physletb.2012.02.013
[arXiv:1112.3022 [hep-ph]].
%
For a more complete set of references see \cite{Markkanen:2018pdo}



\bibitem{Avramidi:1986mj}
I.~G.~Avramidi,
[arXiv:hep-th/9510140 [hep-th]].

\bibitem{1979577}
M.~M.~Defranchis [ATLAS and CMS],
[arXiv:2105.05776 [hep-ex]].


\bibitem{Buchbinder:1992rb}
  I.~L.~Buchbinder, S.~D.~Odintsov and I.~L.~Shapiro,
  Bristol, UK: IOP (1992) 413 p
  
\bibitem{Markkanen:2018pdo}
T.~Markkanen, A.~Rajantie and S.~Stopyra,
Front. Astron. Space Sci. \textbf{5} (2018), 40
doi:10.3389/fspas.2018.00040
[arXiv:1809.06923 [astro-ph.CO]].
  
\bibitem{Markkanen:2018bfx}
T.~Markkanen, S.~Nurmi, A.~Rajantie and S.~Stopyra,
JHEP \textbf{06} (2018), 040
doi:10.1007/JHEP06(2018)040
[arXiv:1804.02020 [hep-ph]].

\bibitem{Figueroa:2017slm}
D.~G.~Figueroa, A.~Rajantie and F.~Torrenti,
Phys. Rev. D \textbf{98} (2018) no.2, 023532
doi:10.1103/PhysRevD.98.023532
[arXiv:1709.00398 [astro-ph.CO]].



\bibitem{Rajantie:2017ajw}
A.~Rajantie and S.~Stopyra,
Phys. Rev. D \textbf{97} (2018) no.2, 025012
doi:10.1103/PhysRevD.97.025012
[arXiv:1707.09175 [hep-th]].

\bibitem{Herranen:2015ima}
M.~Herranen, T.~Markkanen, S.~Nurmi and A.~Rajantie,
Phys. Rev. Lett. \textbf{115} (2015), 241301
doi:10.1103/PhysRevLett.115.241301
[arXiv:1506.04065 [hep-ph]].

\bibitem{Herranen:2014cua}
M.~Herranen, T.~Markkanen, S.~Nurmi and A.~Rajantie,
Phys. Rev. Lett. \textbf{113} (2014) no.21, 211102
doi:10.1103/PhysRevLett.113.211102
[arXiv:1407.3141 [hep-ph]].


\bibitem{Vilenkin:1985md}
A.~Vilenkin,
Phys. Rev. D \textbf{32} (1985), 2511
doi:10.1103/PhysRevD.32.2511

\bibitem{Jedamzik:2010dq}
K.~Jedamzik, M.~Lemoine and J.~Martin,
JCAP \textbf{09} (2010), 034
doi:10.1088/1475-7516/2010/09/034
[arXiv:1002.3039 [astro-ph.CO]].

\bibitem{Lacey:1993iv}
C.~G.~Lacey and S.~Cole,
Mon. Not. Roy. Astron. Soc. \textbf{262} (1993), 627-649

\bibitem{Berezinsky:2003vn}
V.~Berezinsky, V.~Dokuchaev and Y.~Eroshenko,
Phys. Rev. D \textbf{68} (2003), 103003
doi:10.1103/PhysRevD.68.103003
[arXiv:astro-ph/0301551 [astro-ph]].



\bibitem{Planck:2018vyg}
N.~Aghanim \textit{et al.} [Planck],
Astron. Astrophys. \textbf{641} (2020), A6
doi:10.1051/0004-6361/201833910
[arXiv:1807.06209 [astro-ph.CO]];
Planck 2018 explanatory supplement, wiki.cosmos.esa.int/planckpla2018/, “Mission
Products: Cosmological parameters: Parameter Tables”.

\bibitem{Sotiriou}
T.~Sotirious and V.~Faraoni,
Rev. Mod. Phys. \textbf{82} (2010), 451-497
doi:10.1103/RevModPhys.82.451
[arXiv:0805.1729 [gr-qc]].

\bibitem{BarausseSotiriou}
E.~Barausse and T.P.~Sotiriou,
Phys. Rev. Lett. \textbf{101} (2008), 099001
doi:10.1103/PhysRevLett.101.099001
[arXiv:0803.3433 [gr-qc]].

\bibitem{TattersallFerreira}
O.J.~Tattersall and P.G.~Ferreira,
Phys. Rev. D. \textbf{97} (2018), 104047
doi:10.1103/PhysRevD.97.104047
[arXiv:1804.08950 [gr-qc]].

\bibitem{FerreiraTattersall}
P.G.~Ferreira and O.J.~Tattersall,
Phys. Rev. D. \textbf{101} (2020), 024011
doi:10.1103/PhysRevD.101.024011
[arXiv:1910.04480 [gr-qc]].

\bibitem{Gorbunov:2010bn}
D.~Gorbunov and A.~Panin,
Phys. Lett. B \textbf{700} (2011), 157-162
doi:10.1016/j.physletb.2011.04.067
[arXiv:1009.2448 [hep-ph]].

\bibitem{Gorbunov:2012ns}
D.~Gorbunov and A.~Tokareva,
JCAP \textbf{12} (2013), 021
doi:10.1088/1475-7516/2013/12/021
[arXiv:1212.4466 [astro-ph.CO]].

\bibitem{Gorbunov:2013lna}
D.~Gorbunov and A.~Tokareva,
Phys. Part. Nucl. Lett. \textbf{10} (2013), 633-636
doi:10.1134/S1547477113070030





\bibitem{Bernal:2020qyu}
  N.~Bernal, J.~Rubio and H.~Veermae,
  [arXiv:2006.02442 [hep-ph].]
  
  \bibitem{Ghilencea:2021lpa}
D.~M.~Ghilencea,
[arXiv:2104.15118 [hep-ph]];
Phys. Rev. D \textbf{101} (2020) no.4, 045010
doi:10.1103/PhysRevD.101.045010
[arXiv:1904.06596 [hep-th]],
JHEP \textbf{03} (2019), 049
doi:10.1007/JHEP03(2019)049
[arXiv:1812.08613 [hep-th]];


\end{thebibliography}
\end{document}